\documentclass[fleqn,twoside]{article}
\usepackage[headings]{espcrc2}

\readRCS
$Id: bPhysics04.tex,v 1.7 2004/11/01 09:41:57 slava77 Exp $
\usepackage{graphicx}
\usepackage[figuresright]{rotating}

\newcommand{\AmS}{{\protect\the\textfont2
  A\kern-.1667em\lower.5ex\hbox{M}\kern-.125emS}}

\def\ra{\rightarrow}

\newcommand{\ctau}{\mbox{$c\tau\,$}}

\newcommand{\TeV}{\ensuremath{\mathrm{Te\kern -0.1em V}}}
\newcommand{\GeV}{\ensuremath{\mathrm{Ge\kern -0.1em V}}}
\newcommand{\MeV}{\ensuremath{\mathrm{Me\kern -0.1em V}}}
\newcommand{\GeVc}{\ensuremath{\GeV/c}}
\newcommand{\GeVcc}{\ensuremath{\GeV/{c^2}}}

\newcommand{\MeVcc}{\ensuremath{\MeV/{c^2}}}
\newcommand{\Tesla}{\ensuremath{\mathrm{T}}}

\newcommand{\um}{\ensuremath{\mathrm{\mu m}}}

\newcommand{\pbin}{\ensuremath{\mathrm{pb}^{-1}}}
\newcommand{\fbin}{\ensuremath{\mathrm{fb}^{-1}}}

\newcommand{\dz}{\ensuremath{D^0}}
\newcommand{\bs}{\ensuremath{B_s^0}}
\newcommand{\bsd}{\ensuremath{B_{s(d)}^0}}
\newcommand{\bd}{\ensuremath{B_d^0}}
\newcommand{\bp}{\ensuremath{B^+}\,}
\newcommand{\kp}{\ensuremath{K^+}\,}
\newcommand{\jp}{\ensuremath{J/\psi\,}}
\newcommand{\mm}{\ensuremath{\mu^{+}\mu^{-}}}

\newcommand{\pipi}{\ensuremath{\pi^{+}\pi^{-}}}
\newcommand{\pik}{\ensuremath{K^{-}\pi^{+}}}

\newcommand{\jpmm}{\ensuremath{\jp\!\ra\mm\,}}
\newcommand{\bmm}{\ensuremath{B_{s(d)}^0\ra\mm\,}}
\newcommand{\bsmm}{\ensuremath{\bs\ra\mm}}
\newcommand{\bdmm}{\ensuremath{\bd\ra\mm}}
\newcommand{\dzmm}{\ensuremath{\dz\ra\mm}}
\newcommand{\dzpp}{\ensuremath{\dz\ra\pipi}}
\newcommand{\dzpk}{\ensuremath{\dz\ra\pik}}
\newcommand{\bpjpk}{\ensuremath{\bp\ra\jp\kp\,}}
\newcommand{\bpjppi}{\ensuremath{\bp\ra\jp\pi^{+}\,}}
\newcommand{\roots}{\ensuremath{\sqrt{s}}}
\newcommand{\Xp}{\ensuremath{X(3872)}}

\title{$B$-physics: new states, rare decays and branching ratios in CDF}

\author{Vyacheslav Krutelyov,\address[TAMUCDF]{
	Department of Physics, Texas A\&M University,
        College Station, TX 77843-4242, USA.}%
	~for CDF Collaboration.}
       
\runauthor{V.~Krutelyov}

\begin{document}

\begin{abstract}
We present results and prospects for searches for rare $B$ and $D$ meson decays
with final state dimuons, including \bsmm, \bdmm, and \dzmm. Upper limits on the 
branching fractions are compared to previous CDF measurements, recent results from
the $B$ factories and theoretical expectations.
We also report on new measurements of production and decay properties of the 
\Xp\ particle, discovered in 2003 by the Belle Collaboration.
New results on the measurement of the relative branching fraction for the Cabibbo suppressed decay \bpjppi\ 
$\mathcal{B}(\bpjppi)/\mathcal{B}(\bpjpk)$ are presented too.
The presented results are based on the analyses of $70$ to $220~\pbin$ of 
data collected by the CDF~II detector in $p\bar{p}$ collisions at $\roots = 1.96~\GeV$
at Fermilab Tevatron.
\vspace{1pc}
\end{abstract}

\maketitle

\section{INTRODUCTION}
The Tevatron collider at Fermilab continues its operation 
of Run~II since 2001. The $p\bar{p}$ collisions are produced at $\roots = 1.96~\GeV$ with yet
increasing instantaneous luminosity. The amount of integrated luminosity delivered to the CDF and D$\O$
experiments has surpassed that of Run~I by nearly a factor of 5, exceeding $0.5~\fbin$ by July 2004.

The $p\bar{p}$ collisions are an abundant source of Beauty and Charm hadrons. 
The upgraded Collider Detector at Fermilab (CDF~II) ~\cite{CDFtdr} can effectively
select many of the decay modes of the beauty and charm hadrons. This allows for an analysis or search
of statistics limited, rare processes.

\section{DETECTOR}
The components of the CDF~II detector pertinent to the analyses presented here are described briefly below.
Detailed description can be found elsewhere~\cite{CDFtdr}.
The cylindrical drift chamber (COT) and the silicon microstrip detector (SVX~II)
are immersed in a $1.4~\Tesla$ solenoidal magnetic field and allow for tracking charged particles in the range of $|\eta|<1$
with good position precision (impact parameter, $d_0$, resolution $\sigma_{d_0} \sim 15~\um$) 
and good momentum precision (transverse momentum, $p_T$, resolution $\delta p_T/p_T^2 \sim 0.1\%/(\GeVc)$).
The muon subdetector includes the CMU (detectable muon $p_T>1.5~\GeVc$), CMP ($p_T > 2.5~\GeVc$) covering the pseudorapidity
range of $|\eta|<0.6$, and CMX ($p_T>2~\GeVc$) covering the range of $0.55<|\eta|<1.0$.

\subsection{Triggers}
CDF~II employs a 3 level trigger system.
Two sets of triggers are used by the analyses presented here: the {\it dimuon} and {\it two-track} triggers.

The {\it dimuon} trigger selects two muons (track associated with the hits in a muon subdetector)
with $p_T>1.5~\GeVc$.

The {\it two-track} trigger selects tracks with $p_T>2~\GeVc$ and uses the Level-2 Silicon Vertex Tracker (SVT)
to select on the impact parameter of a track with respect to the beamline. The impact parameter resolution is
below $50~\um$.
The tracks are selected with 
$120~\um < |d_0| < 1.0$~mm.

\boldmath
\section{\dzmm\ SEARCH}
\unboldmath
The flavor-changing neutral current (FCNC) decay \dzmm\ is highly suppressed in the Standard Model (SM),
$\mathcal{B}_{SM}(\dzmm)\approx 10^{-13}$~\cite{dzmmPRD}. 
This prediction is many orders of magnitude beyond the reach of the present 
experiments. 
Observation of this decay at a rate significantly
exceeding the SM expectation would indicate the presence of non-SM particles or couplings. 
Thus a large, unexplored region exists in which to search for new physics. 

This search uses a $65~\pbin$ data sample.
The sample of $\sim 10^5$ $D^*$-tagged two-body \dz\ decays selected by the two-track trigger is
used to estimate backgrounds, to optimize selection requirements, and to normalize the sensitivity of the search
from the data sample itself.

The $\mathcal{B}(\dzmm)$ upper limit for the given confidence level (CL) is determined using
\begin{equation}
\label{eqn:dzmm}
\frac{\mathcal{B}_{\mathrm{CL}}(\dzmm)}{\mathcal{B}(\dzpp)} =
\frac{N_{\mathrm{CL}}(\mu\mu)}{N(\pi\pi)} \frac{\epsilon(\mu\mu)}{\epsilon(\pi\pi)} \frac{\alpha(\mu\mu)}{\alpha(\pi\pi)},
\end{equation}
where $\mathcal{B}(\dzpp) = (1.43 \pm 0.07) \times 10^{-3}$ is the 
normalization branching fraction, $N(\pi\pi)$
is the number of \dzpp\ events observed, $N_{\mathrm{CL}}(\mu\mu)$ is the upper limit on the
observed \dzmm\ signal events with the expected background accounted for.
 The $\epsilon$ and $\alpha$ are the efficiency and acceptance
for each mode. 
Except for the 
muon identification, and the assignment of different particle masses, the same selection requirements
are applied to both modes. 
Kinematically,
 the \dzpp\ and \mm\ modes are nearly identical,
minimizing the systematic uncertainty 
and the differences in acceptance.
The efficiency and acceptance fractions are estimated to be $\epsilon(\pi\pi)/\epsilon(\mu\mu)=1.1\pm0.04$ and
$\alpha(\pi\pi)/\alpha(\mu\mu)=0.96\pm0.02$.
The width of the reconstructed mass peak for two-body decays of 
the \dz\ 
is 
sufficient to separate \dzpk\ 
from \pipi.

The dominant sources of the background here are the combinatorial background (estimated using the events in the high mass sideband),
and the muon misidentification background, when a pion is misidentified as a muon (estimated using the sample of $D^*$-tagged 
$D^0\ra K^- \pi^{+}$ events).

A ``blinded'' analysis was performed. 
The data in the signal 
mass window were hidden and the analysis cuts optimized without knowledge of their actual impact on the result.

Using the optimized selection requirements, 
$5.0 \pm 2.2$ events remain in the high mass sideband, yielding $1.6 \pm 0.7$ expected
from the flat component of the background. 
The expected number of misidentification events is $0.22\pm0.02$.
The total expected background is $1.8\pm0.7$ events. The number of events in the normalization mode is $N(\pi\pi) = 1412\pm 54$.
\begin{figure}[thb]
  \includegraphics[width=17pc,height=8pc]{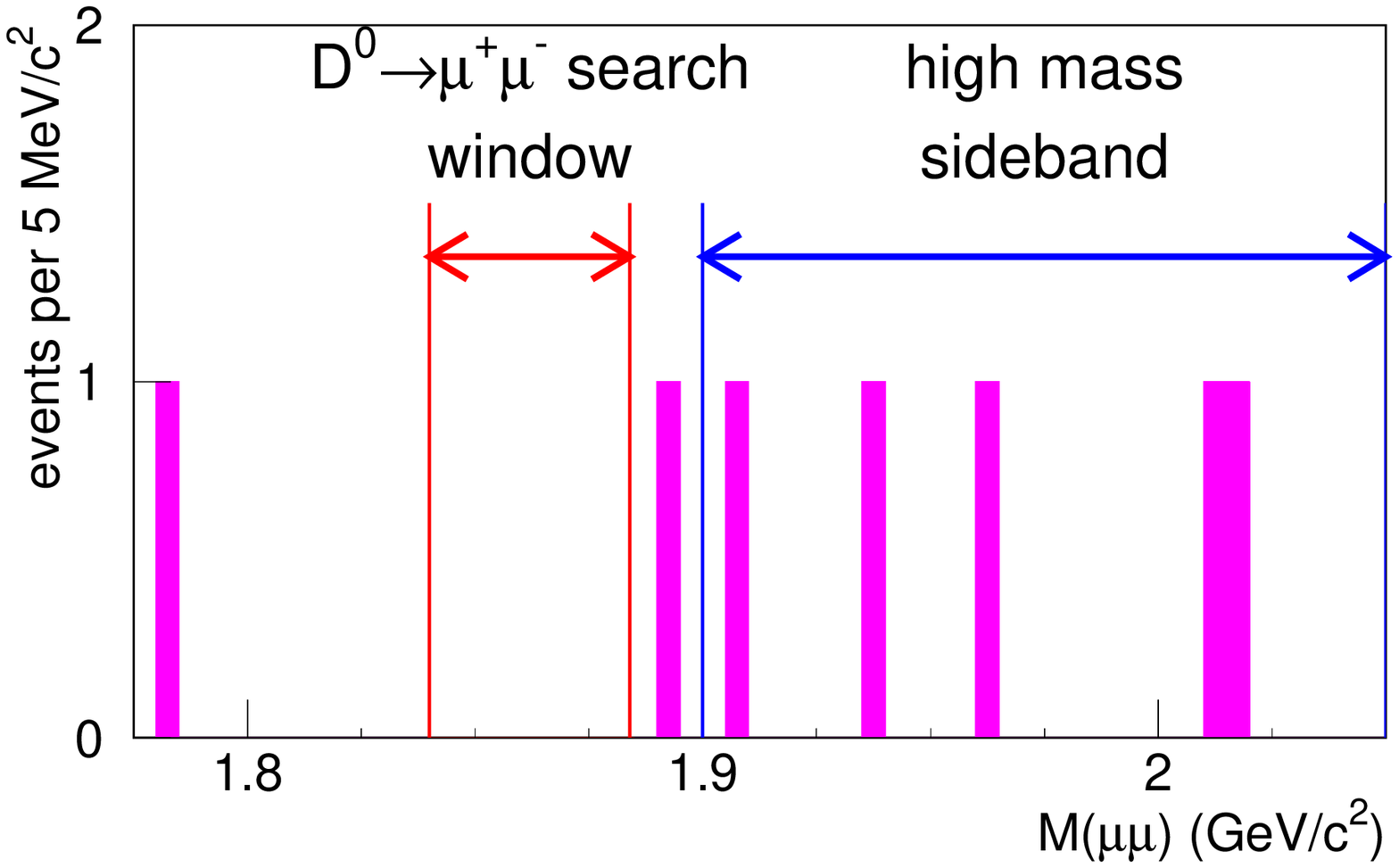}
  \caption{\label{fig:dmmOpenBox}
The mass distribution of candidate \dzmm\ events. 
  }
\end{figure}
We apply the optimized selection requirements to the signal region of the $\mu\mu$ sample and find no events remaining, 
as displayed in Fig.~\ref{fig:dmmOpenBox}. Using Eq.~(\ref{eqn:dzmm}) we find an upper limit on the branching fraction of
\begin{equation}
\mathcal{B}(\dzmm) \leq 2.5 \times 10^{-6} (3.3 \times 10^{-6})
\end{equation} 
at the $90\%$ ($95\%$) confidence level. This result improves on the best previously published limits. This limit has recently been 
improved by HERA-B to the value of $2\times 10^{-6}$ at $90\%$ confidence level~\cite{dzmmHERA}. 
We expect to improve this limit using a substantially larger dataset.

\boldmath
\section{\bmm\ SEARCH}
\unboldmath
The rare FCNC decay \bsmm\ is one of the most sensitive probes to physics beyond the SM~\cite{bmmPRL}. The decay has not been
experimentally observed and the best previously published branching ratio limit
was $\mathcal{B}(\bsmm)<2.0\times 10^{-6}$ at $90\%$ CL, while the SM prediction is $(3.5\pm0.9)\times 10^{-9}$. 
Similarly, the best previously published limit on the related branching ratio, $\mathcal{B}(\bdmm) < 1.6 \times 10^{-7}$,
is about three orders of magnitude larger than its SM expectation. 
The $\mathcal{B}(\bsmm)$ can be enhanced by one to three orders of magnitude
in various supersymmetric (SUSY) extensions of the SM. 

The presented analysis uses $171~\pbin$ of data collected through September 2003. 
The data sample is collected using the dimuon trigger, selecting the muons with $|\eta|<0.6$.

We model the signal \bmm\ decays using the Pythia Monte Carlo (MC).
To normalize to experimentally determined cross-section, we require the \bsd\ to have $p_T(\bsd)>6~\GeVc$ and rapidity
$|y|<1$.

To further discriminate \bmm decays from background events we use the following four variables: the invariant mass
$(M_{\mm})$; the $B$-candidate proper decay length ($\ctau$);
the opening angle ($\Delta\Phi$) between the $B$-hadron
flight direction (estimated as the vector $\vec{p}_T^{\mu\mu}$) 
and the direction to the decay vertex;
and the $B$-candidate track isolation 
($I\equiv p_T^{\mu\mu}/(\sum p_T^{\mathrm{trk}} + p_T^{\mu\mu})$, where 
$\Delta R(p_T^{\mathrm{trk}}, p_T^{\mu\mu}) \equiv\sqrt{(\Delta\phi)^2+(\Delta\eta)^2}<1.0$).

We use a ``blind'' analysis technique.
The data in the search mass window $5.169 < M_{\mu\mu} < 5.469~\GeVcc$ are hidden (corresponding to $\approx\pm 4$
times the mass resolution and is centered on the \bs\ and \bd\ masses).
The optimization performed using only data in the mass sideband regions, 
$M_{\mu\mu}\in[4.669, 5.169]\cup[5.469,5.969]~\GeVcc$. 

We vary among the sets of $(M_{\mu\mu}, \ctau, \Delta\Phi, I)$ criteria  to
minimize the {\it a priori} expected $90\%$~CL upper limit on the branching ratio. 
For a given number of observed 
events, $n$, and an expected background of $n_{\mathrm{bg}}$, the branching ratio is determined using:
\begin{equation}
\label{eqn:bmmLimit}
\mathcal{B}(\bsmm) \leq \frac{N(n, n_{\mathrm{bg}})}{2 \sigma_{\bs} \mathcal{L} \alpha \epsilon_{\mathrm{total}}},
\end{equation}
where $N(n, n_{\mathrm{bg}})$ is the number of candidate \bsmm\ decays at $90\%$~CL,
and $\mathcal{L}$ is the integrated luminosity.
The {\it a priori} 
expected limit is given by the sum over all 
$n$, weighted by the 
Poisson probability
$P(n|n_{\mathrm{bg}})$.
For the \bdmm\ limit we substitute $\sigma_{\bd}$ for $\sigma_{\bs}$.
The factor of two in the denominator accounts for the charge-conjugate $B$-hadron final states.
The total acceptance times efficiency, $\alpha\epsilon_{\mathrm{total}}$, and $n_{\mathrm{bg}}$ 
are estimated separately for each set.

The acceptance is estimated using the MC sample. The total efficiency is a product of reconstruction, trigger, and 
the analysis selection. The reconstruction and trigger efficiencies are estimated using unbiased samples of \jpmm\ decays.
The efficiency of the analysis selection criteria is determined using the MC sample.

\begin{figure}[htb]
  \includegraphics[width=15pc, height=12pc]{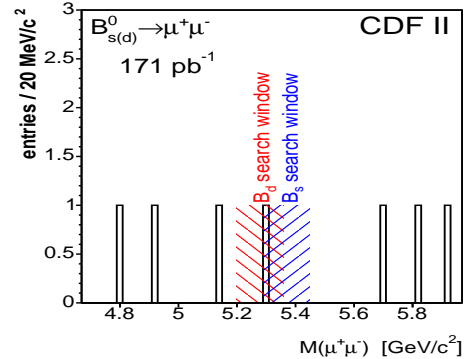}
  \caption{\label{fig:bmmOpenBox}
The \mm invariant mass distribution of the events in the sideband and search regions satisfying
all requirements. 
  }
\end{figure}

The optimization procedure yields $(\ctau >200~\um, \Delta\Phi < 0.1~\mathrm{rad}, I >0.65)$ and the mass window $\pm 80~\MeVcc$
around 
\bs\ (5.369~\GeVcc) and \bd\ (5.279~\GeVcc) masses. The estimated acceptance is $\alpha\approx6.6\%$
and the total efficiency is $\epsilon_{\mathrm{total}}\approx 30\%$. 
The expected background is $1.1\pm0.3$ events in each of the \bs\ and \bd\ 
mass windows. Using the optimized set of selection criteria, one event survives all requirements and has an invariant mass of
$M_{\mu\mu}=5.295~\GeVcc$, thus falling into both the \bs\ and \bd\ search windows as shown in Fig.~\ref{fig:bmmOpenBox}. 
Using Eq.~(\ref{eqn:bmmLimit}) we derive $90\%$ ($95\%$) CL limits of $\mathcal{B}(\bsmm)<5.8 \times 10^{-7} (7.5\times 10^{-7})$
$\mathcal{B}(\bdmm)<1.5 \times 10^{-7} (1.9\times 10^{-7})$. The new limit improves the previously best published limit
by a factor of three and significantly reduces the allowed parameter space of $R$-parity violating and $SO(10)$ 
SUSY models. We expect to improve the limit by using the larger dataset as well as enhancing the
analysis techniques~\cite{bmmUpdates}.

\boldmath
\section{\Xp\ PROPERTIES}
\unboldmath
In 2003 the Belle Collaboration reported a new particle, \Xp, observed in exclusive decays of $B$-mesons produced in 
$e^+e^-$ collisions~\cite{xpartPRL}. This particle has a mass of $3872~\MeVcc$ and decays to \jp\pipi. A natural 
interpretation of this particle would be a previously unobserved charmonium state, but there are no such states predicted
to lie at or near the observed mass with the right quantum numbers to decay into \jp\pipi.
Whether it is a new form of hadronic matter or a conventional $c\bar{c}$-state in conflict with theoretical models,
the \Xp\ is an important object of study. Here we report the observation of a \jp\pipi\ resonance produced inclusively
in $p\bar{p}$ collisions and which is consistent with the \Xp. We also report the measurement of the \Xp\ and $\psi(2S)$ 
production fraction
associated with a large lifetime ($B$-hadrons)~\cite{xpartLifePub}.

The analysis uses $220~\pbin$ of data 
collected using the dimuon trigger selecting the muons associated with CMU or CMX muon chambers within the mass range of
$2.7$ to $4.0$~\GeVcc.
\begin{figure}[htb]
  \includegraphics[width=16pc, height=14pc]{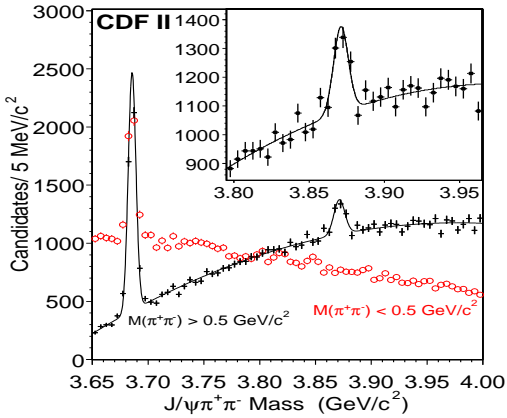}
  \caption{\label{fig:xpMass}
The \jp\pipi\ mass distributions for the sample separated into those candidates with
dipion masses less than (circles) or greater than (solid points) $500~\MeVcc$.
  }
\end{figure}

In the offline analysis, the two opposite charge tracks with $p_{T}>0.4~\GeVcc$ are added to the \jp\ candidate. All four tracks are then 
required to come from the same vertex. Additional strict requirements are made to suppress background: \jp\ mass within 60~\MeVcc 
($\sim 4 \sigma_{m_{\mu\mu}}$), $p_T(\jp)>4~\GeVc$, good vertex probability, and $\Delta R<0.7$.

Besides the 
$\psi(2S)$ peak,  another peak is observed at 
the \jp\pipi\ mass of around $3872.3~\MeVcc$. The Gaussian plus a quadratic polynomial is used to model each peak.
The binned 
likelihood fit results $5790\pm140$ $\psi(2S)$ candidates and $580\pm100$ \Xp\ candidates.

The \Xp\ signal reported by Belle Collaboration favors large \pipi\ masses. Our data supports this conclusion as well, 
see Fig.~\ref{fig:xpMass}. Requiring the $M_{\pipi} > 500~\MeVcc$ reduces the background by almost a factor of two. 
We use the high mass sample for measuring the \Xp\ mass, we find $m(\Xp) = 3871.3\pm0.7(stat)\pm0.4(syst)~\MeVcc$. This is
in agreement with the measurement by the Belle Collaboration. The observed width of $4.2\pm0.8~\MeVcc$ is found to be consistent with 
the detector mass resolution.

For the lifetime analysis the data sample is the same as for the mass measurement.
The $M_{\pipi}>500~\MeVcc$ is applied to the \Xp\ signal region as it greatly improves the purity.
This cut is not applied for the $\psi(2S)$ analysis, as it has a strong effect on the number of signal 
events due to the lower mass of the $\psi(2S)$.
\begin{figure}[ht]
  \includegraphics[width=17pc]{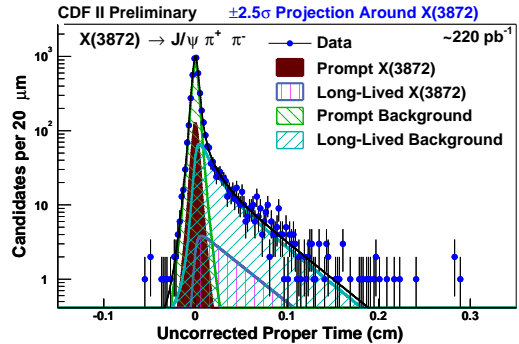}
  \caption{\label{fig:xpSBSfraction}
   The uncorrected proper-time 
projection of the full \Xp\ likelihood PDF over the restricted mass range $\pm2.5$ standard deviations around the \Xp\ mass.
  }
\end{figure}

To separate the prompt from long-lived components we perform an unbinned likelihood fit in which 
mass and lifetime 
are simultaneously included. Each candidate is characterized by its mass $M$, uncorrected
proper time, $c\tau \equiv \frac{M(\jp\pi\pi)}{p_T(\jp\pi\pi)}L_{\mathrm{xy}}$, 
and the uncertainty on $c\tau$, $\sigma_{c\tau}$. A projection of the fit for \Xp\ is shown in Fig.~\ref{fig:xpSBSfraction}.

We find the long-lived fractions of the sample to be: $F(\psi(2S)) = 28.3 \pm1.0(stat)\pm0.7(syst)\%$ 
and $F(\Xp) = 16.1\pm4.9(stat)\pm2.0(syst)\%$. The lifetimes of the long-lived signal components are
consistent with the latter being produced in the $B$-hadron decays.

\boldmath
\section{\bpjppi\ MEASUREMENT}
\unboldmath
The Cabibbo-suppressed decay \bpjppi proceeds via a $b\ra (c\bar{c}) +  d$ transition. 
The modes governed by this transition may show direct CP violating effects at the few percent level~\cite{babarBpjppi2004}.
The \bpjppi\ is expected to have a rate about $5\%$ of
that of the Cabibbo-allowed mode \bpjpk. 
The PDG 2004 average
of the ratio of branching ratios is $4.0\pm0.5\%$~\cite{pdg2004}, and the more recent result from BABAR is 
$5.37\pm0.37(stat)\pm0.11\%$~\cite{babarBpjppi2004}.
We present the measurement of the ratio of branching ratios 
$\mathcal{B}_{\mathrm{rel}} \equiv \mathcal{B}(\bpjppi)/\mathcal{B}(\bpjpk)$ using the CDF~II data.

Distinguishing $\jp\pi^+$ from \jp\kp\ events and determining the signal significance is a complicated task since the two modes 
overlap kinematically. The solution is to fit in $\jp\pi^+$ and \jp\kp\ mass space simultaneously.
We use the following relation to measure the desired value:
\begin{eqnarray*}
\mathcal{B}_{\mathrm{rel}} = \frac{N(\jp\pi^{+})}{N(\jp\kp)} \times \frac{\epsilon_{\jp\kp}}{\epsilon_{\jp\pi^+}} 
= {r_{\mathrm{obs}}}\times\frac{1}{\epsilon_{\mathrm{rel}}},
\end{eqnarray*}
where $r_{\mathrm{obs}}$ is the observed raw ratio of branching 
ratios in data and $\epsilon_{\mathrm{rel}}$ is the relative efficiency of the two decay modes.

The analysis uses $200~\pbin$ of data collected using the dimuon trigger. The third track ($K$ or $\pi$) is required to have
$p_T > 2.0~\GeVc$ and to be consistent to come from the same vertex as the \jp\ candidate. The \bp\ candidate should have
$p_T>6.5~\GeVc$ and have a displaced decay vertex, $L_{\mathrm{xy}}>200~\um$. 
The unbinned likelihood fit is used to extract the sample composition.
The $\jp\pi^+$ and \jp\kp\ are each modeled with a Gaussian, the background in \jp\kp\ is modeled with a first order polynomial. The fit
gives $r_{\mathrm{obs}} = 0.045\pm0.008$ with $N(\jp\kp)=1986\pm36$ and $N(\jp\pi^+)=90\pm15$. 
The study of the systematic uncertainty shows that it is dominated by the signal model.

The relative efficiency is measured using MC with full detector simulation, it is estimated to be 
$\epsilon_{\mathrm{rel}} = 0.991\pm0.008$.

The measured ratio of branching ratios is
$\mathcal{B}_{\mathrm{rel}} = 4.5\pm0.8(stat)\pm0.3(syst)\%$.

\end{document}